\documentclass[acmtog, authorversion]{acmart}

\AtBeginDocument{%
  \providecommand\BibTeX{{%
    \normalfont B\kern-0.5em{\scshape i\kern-0.25em b}\kern-0.8em\TeX}}}

\setcopyright{cc}
\setcctype{by}
\acmJournal{TOG}
\acmYear{2025} \acmVolume{44} \acmNumber{4} \acmArticle{} \acmMonth{8} \acmPrice{}\acmDOI{10.1145/3731429}

\acmSubmissionID{843}

\begin{document}

\title{Light Pipe Holographic Display: Bandwidth-preserved Kaleidoscopic Guiding for AR Glasses}

\author{Minseok Chae}
\email{mschae3d@gmail.com}
\affiliation{%
  \institution{Seoul National University}
  \city{Seoul}
  \country{South Korea}
  \postcode{08826}
}

\author{Chun Chen}
\email{chenchun@snu.ac.kr}
\affiliation{%
  \institution{Seoul National University}
  \city{Seoul}
  \country{South Korea}
  \postcode{08826}
}

\author{Seung-Woo Nam}
\email{711asd@snu.ac.kr}
\affiliation{%
  \institution{Seoul National University}
  \city{Seoul}
  \country{South Korea}
  \postcode{08826}
}

\author{Yoonchan Jeong}
\email{yoonchan@snu.ac.kr}
\affiliation{%
  \institution{Seoul National University}
  \city{Seoul}
  \country{South Korea}
  \postcode{08826}
}

\renewcommand{\shortauthors}{Minseok Chae, et al.}

%%%%%%%%%%%%%%%%%%%%%%%%%%%%%%%%%%%%%%%%%%%%%%%%%%%%%%%%%%%%%%%%%%%%%%%%%%%%%%%%%%%%%%%%%%%%%%%%%%%%%%%%%
%%%%%%%%%%%%%%%%%%%%%%%%%%%%%%%%%%%%%%%%%%%%%%%%%%%%%%%%%%%%%%%%%%%%%%%%%%%%%%%%%%%%%%%%%%%%%%%%%%%%%%%%%


\begin{abstract}
  In this paper, we present a holographic display using a light pipe for augmented reality, and the hologram rendering method via bandwidth-preserved kaleidoscopic guiding method. Conventional augmented reality displays typically share optical architectures where the light engine and image combiner are adjacent. Minimizing the size of both components is highly challenging, and most commercial and research prototypes of augmented reality displays are bulky, front-heavy and sight-obstructing. Here, we propose the use of light pipe to decouple and spatially reposition the light engine from the image combiner, enabling a pragmatic glasses-type design. Through total internal reflection, light pipes have an advantage in guiding the full angular bandwidth regardless of its length. By modeling such kaleidoscopic guiding of the wavefront inside the light pipe and applying it to holographic image generation, we successfully separate the light engine from the image combiner, making the front of the device clear and lightweight. We experimentally validate that the proposed light pipe system delivers virtual images with high-quality and 3D depth cues. We further present a method to simulate and compensate for light pipe misalignment, enhancing the robustness and practicality of the proposed system.
\end{abstract}

%%%%%%%%%%%%%%%%%%%%%%%%%%%%%%%%%%%%%%%%%%%%%%%%%%%%%%%%%%%%%%%%%%%%%%%%%%%%%%%%%%%%%%%%%%%%%%%%%%%%%%%%%
%%%%%%%%%%%%%%%%%%%%%%%%%%%%%%%%%%%%%%%%%%%%%%%%%%%%%%%%%%%%%%%%%%%%%%%%%%%%%%%%%%%%%%%%%%%%%%%%%%%%%%%%%





\begin{teaserfigure}
  \includegraphics[width=\textwidth]{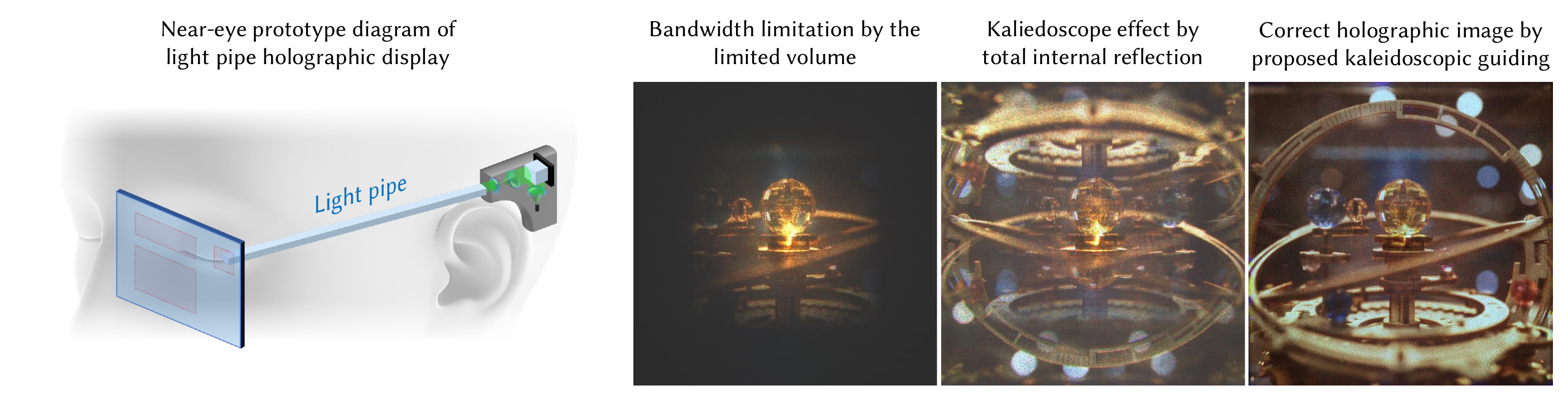}
  \caption{(a) Schematic diagram of the near-eye prototype using the proposed light pipe holographic display. Light engine part including the light source, spatial light modulator, and relay optics are separated from the light guide, leaving the transparent image combiner only in front of the device. (b) Captured holographic image with the limited bandwidth transferred through constrained volume of 3 mm $\times$ 3 mm $\times$ 80 mm, which corresponds to the light pipe volume. (c) Kaleidoscope effect when the holographic image is rendered without consideration of propagation within the light pipe. (d) Correct, full-bandwidth holographic image captured through optimization based on the light pipe's kaleidoscopic guiding model.}
  \label{fig:teaser}
\end{teaserfigure}

\maketitle

%%%%%%%%%%%%%%%%%%%%%%%%%%%%%%%%%%%%%%%%%%%%%%%%%%%%%%%%%%%%%%%%%%%%%%%%%%%%%%%%%%%%%%%%%%%%%%%%%%%%%%%%%
%%%%%%%%%%%%%%%%%%%%%%%%%%%%%%%%%%%%%%%%%%%%%%%%%%%%%%%%%%%%%%%%%%%%%%%%%%%%%%%%%%%%%%%%%%%%%%%%%%%%%%%%%
\section{Introduction} \label{Introduction}

\begin{figure*}[!t]
\centering
 \includegraphics[width=0.95\linewidth]{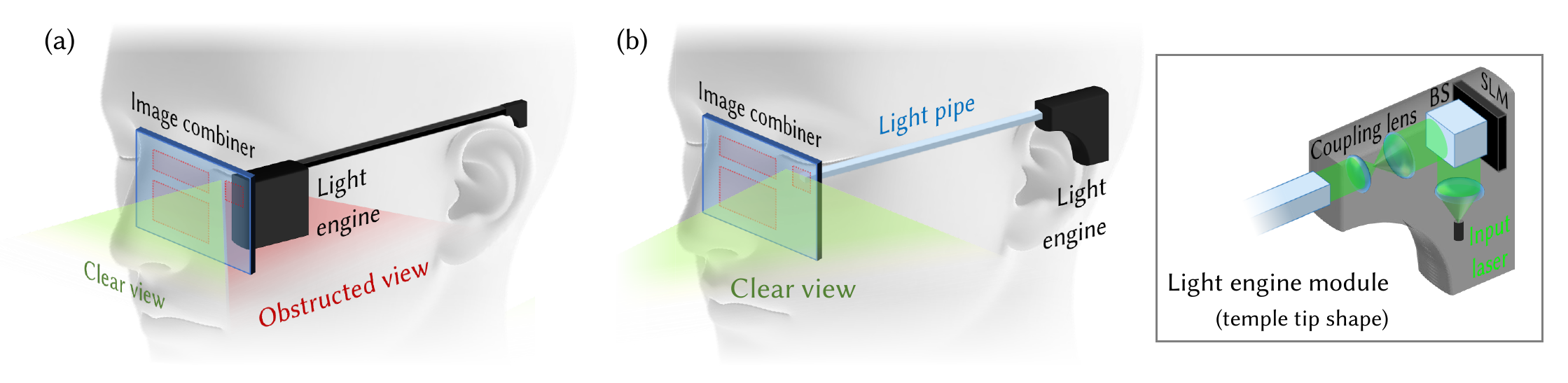}
\caption{Schematic diagram of (a) conventional AR NED structure and (b) the proposed system using light pipe. (a) In conventional NEDs, the light engine is directly attached to the image combiner and obstructs the user's view. Concentration of size and weight at the front of the device results in degraded wearing comfort. (b) The proposed system repositions the light engine away from the image combiner, and distributes the overall weight and form factor. Light pipe guides the wavefront from the engine through total internal reflections and conserves the full bandwidth. Only the transparent image combiner remains at the front, realizing the true glasses-looking device with minimized obstruction.}
\label{fig:necessity}
\end{figure*}

% 홀로그래픽 디스플레이 소개, 홀로그래픽 디스플레이가 near eye display 에서 가지는 장점 소개 : 3차원 표현 가능, wavefront 변조가 가능하기 때문에 광학계 설계 자유도가 높음.
Over the past few decades, significant advances have been made in near-eye displays (NEDs) for augmented reality (AR) across diverse fields such as healthcare, entertainment, and artificial intelligence. AR NEDs enable users to experience unprecedented, immersive interaction between real and virtual objects. AR NEDs can be divided into two main components: the light engine, which consists of the light source, display, and relay optics, and the image combiner, which is positioned in front of the user's eyes. The light engine generates virtual images and floats them at a distant depth. The light emitted from the engine passes through the relay optics and reaches the image combiner, where it is directed into the user’s eyes via a half-mirror or diffractive optical element. Simultaneously, light from the real scene passes through the transparent image combiner, allowing the user to experience the AR scene by combining real and virtual objects.

Conventional AR NEDs are typically designed with the light engine directly attached to the image combiner. This is to ensure that the light from the engine is fully transferred to the combiner. However, since the image combiner is positioned directly in front of the user's eyes, the light engine must also be located near or adjacent to the user's eyes. This configuration increases the weight and bulkiness of the front part of the AR NED. Also, the light engine obstructs the user's field of view (FoV), making it challenging to achieve a seamless AR experience. Although devices with smaller light engine and image combiner have been developed, the inherent limitations of traditional NEDs induce significant challenges to further system miniaturization.

To alleviate such structural challenges, holographic displays can be introduced to AR NEDs. Unlike conventional displays which emit incoherent light, holographic displays are driven by spatial light modulators (SLMs) which are capable of modulating coherent light into arbitrary wavefront at the pixel level [Yaracs et al. \citeyear{yaracs2010state}; Chang et al. \citeyear{chang2020toward}; Shi et al. \citeyear{shi2021towards}]. Through the wavefront modulation, light can be directed in the desired direction or the phase profile of optical components can be reproduced, such as lenses. This distinct property of holographic display offers a higher degree of freedom in the optical system design, and allows the realization of complex optical configurations that are unachievable with incoherent light-based displays. Additionally, holographic displays also have the advantage of providing natural depth cues even for monocular view. Unlike traditional AR devices which rely primarily on binocular disparity to create 3D viewing experience, holographic displays can reconstruct the wavefront as it is propagates from a virtual 3D object. Those unique abilities are particularly advantageous in eyeglasses-type AR NEDs, which require intricate optical systems and natural depth cues to deliver an immersive AR experience.

% 홀로그래픽 NED는 크게 light engine 파트와 image combiner 파트로 이루어져 있음. 근데 홀로그래픽 디스플레이의 특성상 light engine을 이루고 있는 SLM, 광원부, spatial filter 등 차지하는 부피가 많은데, 현재 AR NED는 이것들이 모두 앞에 붙어있어 실용적인 사용이 어려움 보이기
While several holographic AR NEDs have been demonstrated with the light engine integrated into the image combiner, such configurations often involve disadvantages in terms of view obstruction, visual comfort, and weight distribution. A typical holographic light engine consists of a laser light source, collimation optics, a beam splitter to direct planar waves towards the SLM, and a spatial filter to suppress high-order noise induced by the SLM’s pixel structure. These components, along with the necessary electronics and power supply, can contribute to a bulky front-end configuration and view obstruction.
%[Maimone et al. \citeyear{maimone2017holographic}; Gopakumar et al. \citeyear{gopakumar2024full}]

% Image combiner는 무조건 눈앞에 있어야 하기 때문에, 진짜 안경 타입을 만들기 위해서는 light engine 파트를 다른 위치로 옮기고, wavefront만 전달할 수 있는 방법이 좋다.
To achieve obstruction-free and immersive AR NEDs, it is important to minimize the number of optical and electrical components in front of the device. Specifically, the light engine needs to be repositioned, leaving the transparent image combiner only in the user's view. This reposition also has advantages in distributing the weight and form factor to enhance comfortable user experience. Subsequently, a reliable method for guiding the wavefront to the image combiner intact is required, since the light emitted from the engine diverges and loses its bandwidth during propagation.

In this study, we propose the holographic display for AR NED using light pipes as a solution to guide the full bandwidth of the wavefront to the image combiners. Figure \ref{fig:necessity} shows comparison between the conventional NED structure and the proposed system. By utilizing the total internal reflection (TIR) of the light pipe to preserve the whole bandwidth regardless of the length, we successfully reposition the light engine from the image combiner, and demonstrate that the proposed method reconstructs desired images with full bandwidth in AR NEDs. We introduce a propagation model for light pipes tailored for holographic displays, and we develop an algorithm for optimizing the phase profile of SLMs. This model includes a comprehensive methodology for compensating misalignments of the light pipe to make the proposed system more robust. We experimentally demonstrate the ability to reconstruct desired images at multiple depths in AR scene. The contributions of this research are summarized as follows:
\begin{itemize}
  \item Wavefront guiding method using a light pipe to reposition the light engine of AR NEDs and reconstruct desired holographic images.
  \item Analysis and compensation strategies of physical misalignment for system robustness.
  \item Fabrication of the experimental prototypes with the proposed light pipe holography method, and evaluation of the bandwidth conservation.
\end{itemize}

%%%%%%%%%%%%%%%%%%%%%%%%%%%%%%%%%%%%%%%%%%%%%%%%%%%%%%%%%%%%%%%%%%%%%%%%%%%%%%%%%%%%%%%%%%%%%%%%%%%%%%%%%
%%%%%%%%%%%%%%%%%%%%%%%%%%%%%%%%%%%%%%%%%%%%%%%%%%%%%%%%%%%%%%%%%%%%%%%%%%%%%%%%%%%%%%%%%%%%%%%%%%%%%%%%%

\section{Related works} \label{Related works}
\subsubsection*{\indent Holographic AR Displays}

The light engine in conventional AR displays utilizes an incoherent light source or a display unit with relay optics [Kress \citeyear{kress2020optical}; Hsiang et al. \citeyear{hsiang2022ar}; Yin et al. \citeyear{yin2022advanced}]. While incoherent light engines offer the advantages of high image quality and wide field of view (FoV), they face notable challenges, including lack of flexibility in system design, vergence-accommodation conflict (VAC), and a lack of depth cues. To address these limitations, holographic displays have been introduced into AR displays. In holographic AR displays, spatial light modulators such as liquid crystal on silicon (LCoS) [Bleha et al. \citeyear{bleha2013advances}] or digital micromirror devices (DMDs) [Lee et al. \citeyear{lee2022high}], are integrated into the light engine. Recent studies have demonstrated their capability to reconstruct the wavelength with low error [Chakravarthula et al. \citeyear{chakravarthula2019wirtinger}], or reproduce continuous depth cues and full parallax while inherently avoiding the VAC issue [He et al. \citeyear{he2019progress}; Park et al. \citeyear{park2022holographic}; Kim et al. \citeyear{kim2022accommodative}; Lee et al. \citeyear{lee2022high}; Kim et al. \citeyear{kim2024holographic}]. Techniques such as multi-depth and light-field-based hologram optimization [Padmanaban et al. \citeyear{padmanaban2019holographic}; Chen et al. \citeyear{chen2021multi}; Chakravarthula et al. \citeyear{chakravarthula2022hogel}; Schiffers et al. \citeyear{schiffers2023stochastic}], time-multiplexed or polarization-based methods [Makowski et al. \citeyear{makowski2012simple}; Choi et al. \citeyear{choi2022time}; Nam et al. \citeyear{nam2023depolarized}] have been employed to enhance depth cues and parallax realism. Other efforts have focused on extending optical performance [Maimone et al. \citeyear{maimone2017holographic}], including aberration correction [Kim et al. \citeyear{kim2021vision}; Chen et al. \citeyear{chen2022off}; Nam et al. \citeyear{nam2022accelerating}] and expanding spatial bandwidth and eye box size [Jang et al. \citeyear{jang2018holographic}; Lee et al. \citeyear{lee2020wide}; Kuo et al. \citeyear{kuo2020high}; Chae et al. \citeyear{chae2023etendue}; Tseng et al. \citeyear{tseng2024neural}; Chao et al. \citeyear{chao2024large}]. Overall, The advances in hologram optimization algorithms and flexibility in system design make holographic AR display a promising approach for fulfilling the complex demands of human-centric AR experiences in NEDs [Chang et al. \citeyear{chang2020toward}; Blanche et al. \citeyear{blanche2021holography}].

\subsubsection*{\indent Propagation Modeling for Holographic Displays}

Holographic displays offer significant advantages as aforementioned, but these benefits can only be fully realized under ideal optical conditions. In practice, imperfections in optics, such as speckle noise, aberrations, and distortions, create challenges to achieving high-quality holographic images. To develop the improved propagation models of light, researchers have collected computer-generated hologram (CGH) datasets with corresponding camera-captured images [Chakravarthula et al. \citeyear{chakravarthula2020learned}; Choi et al. \citeyear{choi2021neural}; \citeyear{choi2022time}; Yoo et al. \citeyear{yoo2022learning}; Wang et al. \citeyear{wang2024liquid}]. These datasets are used to train parameterized propagation models or neural networks that can realistically simulate light propagation. Another promising approach involves using camera-captured images as feedback in hologram optimization. This feedback provides realistic information, enabling holograms to be updated to correct system imperfections such as speckle noise, structural noise, and inaccurate color reproduction [Peng et al. \citeyear{peng2020neural}; Chen et al. \citeyear{chen2022speckle}; Chen et al. \citeyear{chen2024ultrahigh}]. These developments mark an important step toward narrowing the gap between theoretical and practical applications. This study also apply some of those delicate methods to the optimization process, and provide the accuracy and validity of the proposed light pipe propagation model in Sections \ref{Experiments} and \ref{misalignment_analysis}.

\subsubsection*{\indent Pragmatic System Design}

In recent years, various studies have proposed holographic display systems in reduced size by utilizing its flexibility in system design. This design flexibility allows for customized solutions that address strict requirements in light engines and optical combiners, such as extended eye-box sizes [Shin et al. \citeyear{shin2020eye}; Xia et al. \citeyear{xia2020towards}; Kim et al. \citeyear{kim2022holographic}; Ni et al. \citeyear{ni2023design}], FoV [Lee et al. \citeyear{lee2018metasurface}; Shi et al. \citeyear{shi2018wide}], and full-color capabilities [Li et al. \citeyear{li2021meta}; Gopakumar et al. \citeyear{gopakumar2024full}]. Optical components such as spatial filter or relay optics can be substituted by compact holographic components [Bang et al. \citeyear{bang2019compact}], or even can be omitted by modeling high-order artifacts [Gopakumar et al. \citeyear{gopakumar2021unfiltered}]. More recently, Jang et al. [\citeyear{jang2024waveguide}] modeled the wavefront propagation within the waveguide to simulate and reconstruct the 3D holographic image using thin, compact waveguide image combiner. When combined with those recent works on optical designs, we expect that the proposed light pipe holographic display can offer a promising, comprehensive pathway to realize compact AR NEDs with immersive 3D display performance.

%% ===============================================================================

\section{Propagation models} \label{Propagation models}
%% 수식들 용어 통일하기

\begin{figure}[!t]
\centering
 \includegraphics[width=0.8\linewidth]{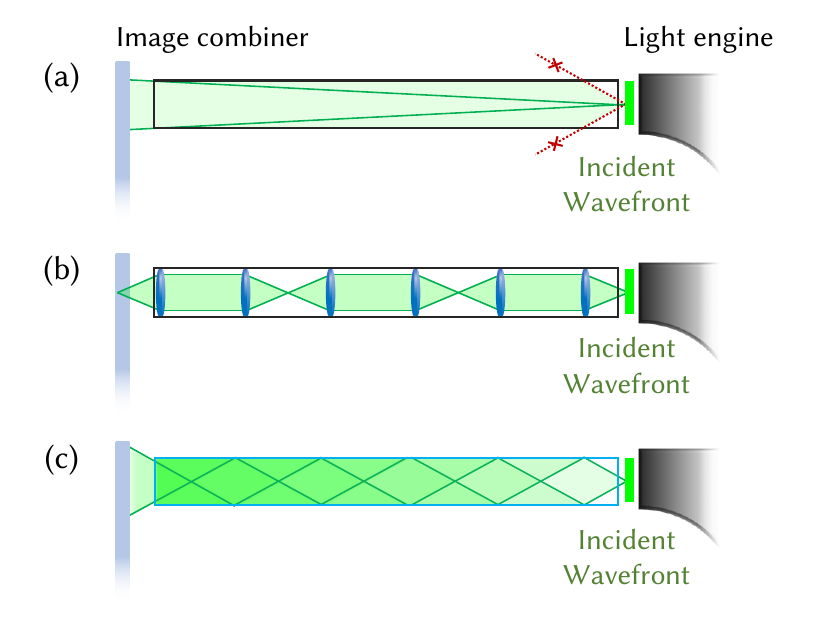}
\caption{Methods for transferring wavefront from the light engine to the image combiner. (a) Free space propagation. The limited volume restricts the transferable bandwidth. (b) 4\textit{f}-system concatenation. (c) The proposed light pipe guiding.}
\label{fig:methods}
\end{figure}

When the light engine is separated from the image combiner, the most essential consideration is to determine how to deliver the wavefront emitted by the light engine to the image combiner. In this chapter, three different wavefront delivery methods are considered: Free space propagation, 4\textit{f}-system concatenation, and the light pipe guiding. Figure \ref{fig:methods} shows the schematic comparison between the three methods.

\subsection{Conventional wavefront transfer methods} \label{Wavefront transfer methods}
%% ASM 같은 기존 propagation 소개
%% NED와 같은 제한된 공간에서 Full bandwidth 전달하려면 공간이 더 필요하다 설명 (시뮬 결과도?)
%% 4f system relay로 전달할 수도 있으나 스펙제한 + 수차 등등 설명
%% 따라서 light pipe를 써서 전달해야 하고, 그에 맞는 모델이 필요하다 피력.
\subsubsection*{\indent Free space propagation}
From the light engine, the wavefront spreads out at a diffraction angle determined by the pixel pitch of the SLM and the wavelength $\lambda_0$ of the input light. The following equations present the maximum diffraction angle $\theta_{\text{max}}$ and the illuminated area $A_{max}$ after propagation within the medium of refractive index $n$:

\begin{equation}
\begin{aligned}
\lambda &= \lambda_0 / n, \\
\theta_{\text{max}} &= \sin^{-1}\left( \frac{\lambda}{2 p_{SLM}} \right), \\
A_{max} &= N p_{SLM} + 2 z \tan\theta_{\text{max}}. \label{eq:max_angle_and_A}
\end{aligned}
\end{equation}

Here, $N$ and $p_{SLM}$ denote the sampling number (along single axis) and sampling pitch of the SLM, respectively. For $N$ = 1000, $p_{SLM}$ = 3.74 $\mu$m, $\lambda_0$ = 638 nm, and $z$ = 80 mm (a specification used in subsequent experiments), the wavefront reaches an area 21.6 times larger than the original SLM area. As $p_{SLM}$ or $z$ increases, the illuminated area expands quadratically. Consequently, the amount of optical information that can reach the identical sampling window decreases significantly, as shown in Fig. \ref{fig:methods} (a). To achieve full angular bandwidth transfer, a physical aperture size of $A_{max}$ is required. This increase in system bulkiness creates challenges for applications in NEDs, where the form factor is highly constrained. 

\begin{figure}[!t]
\centering
 \includegraphics[width=0.88\linewidth]{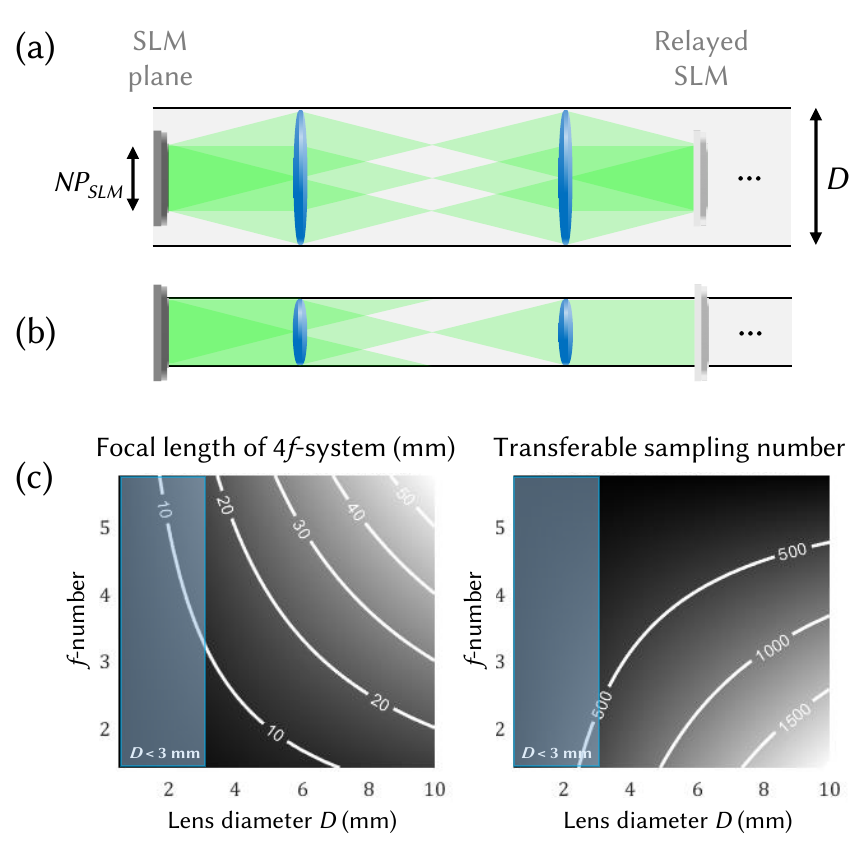}
\caption{The drawback of 4\textit{f}-system when used for wavefront transfer. (a) Required volume for transferring full angular bandwidth. (b) Band-limitation by volume constraint. (c) Lens focal length \textit{f} and transferable SLM sampling number $N$ for 4\textit{f}-system along lens diameter $D$ and lens \textit{f}-number. Blue region indicates the condition where $D < 3 $mm. $p_{SLM}$ and $\lambda_0$ are 3.74 $\mu$m and 638 nm, respectively.}
\label{fig:4fsystem}
\end{figure}

\subsubsection*{\indent 4\textit{f}-system concatenation}
A method widely used for wavefront transfer within limited volume is the 4\textit{f}-system. This system utilizes the Fourier transform properties of lenses to relay the wavefront located at the front focal plane of the first lens to the back focal plane of the second lens. However, the transmittable angular bandwidth is constrained by the \textit{f}-number of lenses. The minimum lens diameter must satisfy the following condition to transfer the full bandwidth:

\begin{equation}
\begin{aligned}
D &> N p_{SLM} + \frac{\lambda_0 f}{p_{SLM}}, \\
N &< \frac{1}{p_{SLM}} \left( D - \frac{\lambda_0 f}{p_{SLM}} \right)
\label{eq:4f_condition1}
\end{aligned}
\end{equation}

% Eq. \ref{eq:4f_condition1} should have solution with feasible $p_{SLM}$ value. Consequently, the following condition must also be fulfilled:

% \begin{align}
% D^2 - 4 N f \lambda &> 0, \quad f < \frac{D^2}{4 N \lambda} \label{eq:4f_condition2}
% \end{align}

This implies that the transferable sampling number of the SLM is limited by the lens specification, as shown in Fig. \ref{fig:4fsystem}. To maximize $N$, the focal length \textit{f} should be minimized with the fixed diameter $D$. Therefore, to transfer the wavefront over long distances, multiple 4\textit{f}-systems must be concatenated. This concatenation accumulates the optical aberrations and geometric distortions by non-ideal lenses, necessitating intricate correction methods to relay the original wavefront intact. Thus, a distinct wavefront transfer method is required for building NED system in more feasible way. In this study, we introduce a light pipe to address those issues and transfer the full bandwidth. Although light pipes may also exhibit aberrations and distortions due to geometric and optical imperfections, high-quality holographic images and simulation-consistent results can be achieved with comparably less correction when the light pipe is manufactured with high geometric precision (further presented in Chapter \ref{Benchtop prototype} and \ref{misalignment_analysis}).

\subsection{Light pipe propagation model} \label{Kaleidoscope}
%% Kaleidoscope 효과 설명,
%% 홀로그램으로는 올바른 이미지 띄우기 가능하다 설명
%% matsushima의 shifted asm 써서 효율적으로 전파모델 가능하다고 설명.

A light pipe is a glass rod made of a uniform medium, commonly used for guiding and distributing or homogenizing the light to enhance the versatility and uniformity of light sources. Through total internal reflection (TIR), the full bandwidth of the wavefront can be transmitted from one end of the light pipe to the other, enabling the reconstruction of a whole holographic image. However, directly applying a free-space propagation model to render the CGH results in incorrect images, where reflected terms are overlapped and create a kaleidoscopic effect. To prevent this, a propagation model that accounts for the kaleidoscope effect must be used for CGH rendering.

\begin{figure}[!t]
\centering
 \includegraphics[width=\linewidth]{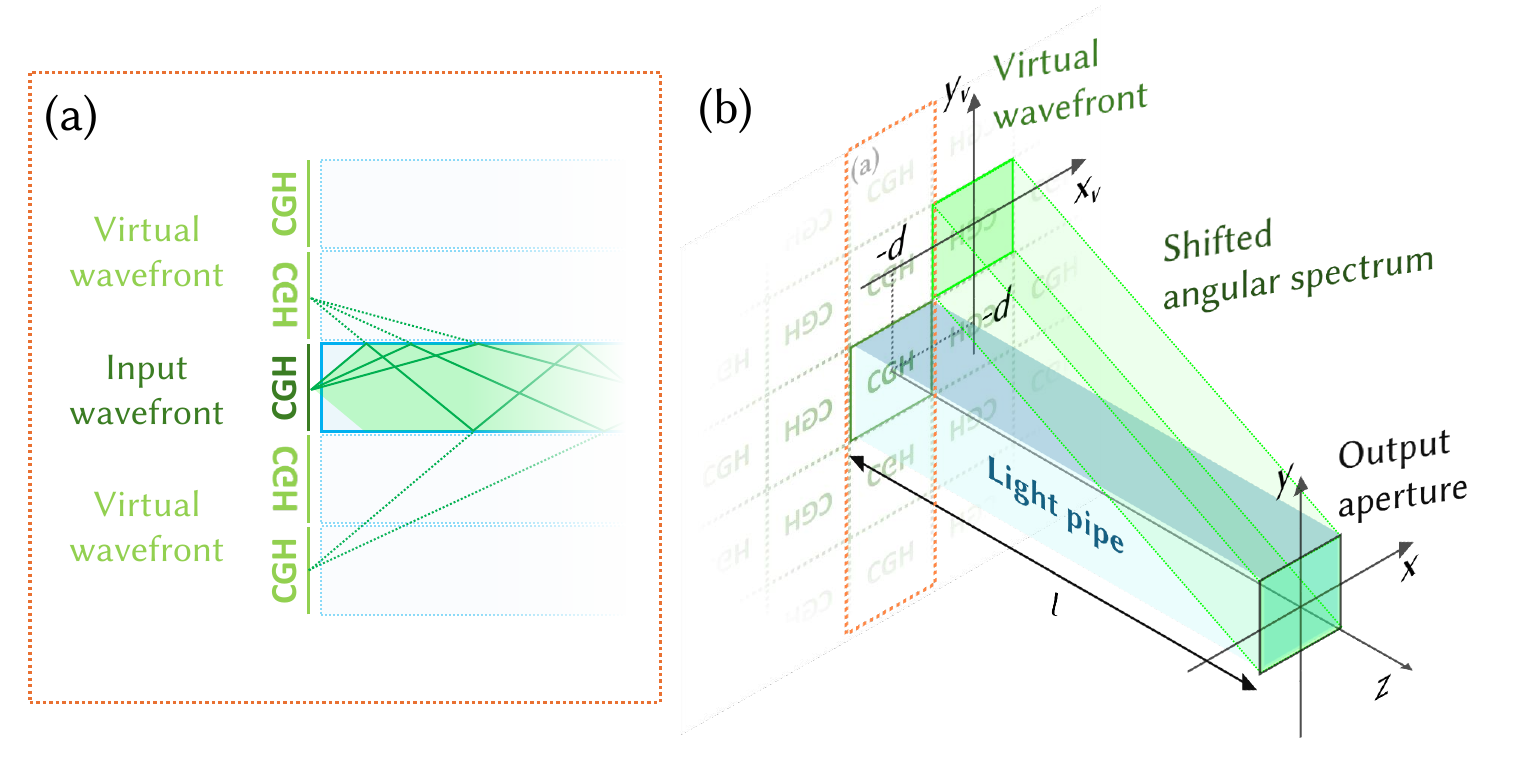}
\caption{Schematic diagram of propagation within the light pipe. (a) The formation of virtual wavefront induced by TIR. (b) The shifted angular spectrum of the virtual wavefront toward the light pipe's output aperture. The orange rectangle indicates the region of (a).}
\label{fig:LP_propagate}
\end{figure}

\begin{figure*}[!t]
\centering
 \includegraphics[width=\linewidth]{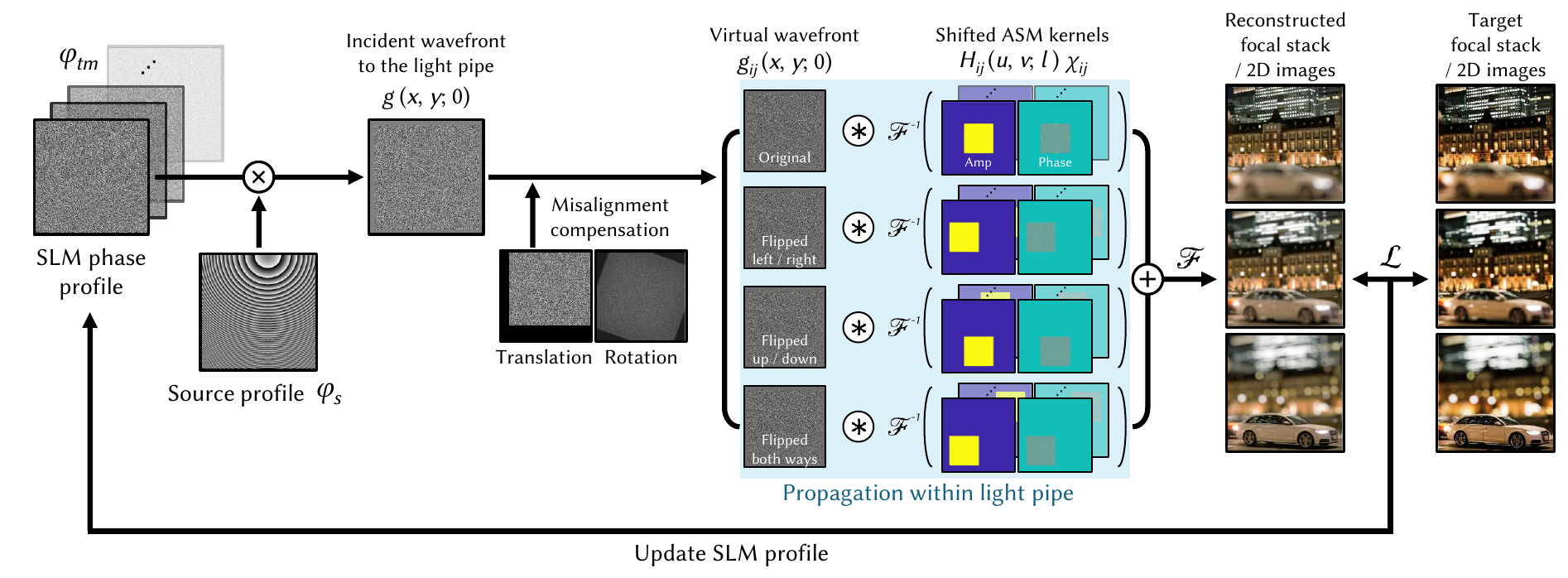}
\caption{SLM phase profile optimization process through light pipe propagation. Temporal multiplexing can be applied to reduce the speckle noise. Source profile is multiplied to the SLM phase profile to remove the DC noise of Fourier holograms (see Section \ref{Experiments}). After correcting the misalignment of the light pipe, the pre-calculated shifted ASM kernels are convolved with the original or flipped virtual wavefronts. Note that shifted ASM kernels are illustrated in Fourier domain to clarify the distinct kernels for each virtual wavefront. By summing up the convolution results, the wavefront at the output aperture of the light pipe can be obtained. Then, the SLM profile is updated based on the loss between target and reconstructed intensities.}
\label{fig:optimization}
\end{figure*}

Research on light behavior within light pipes has been conducted in the field of illumination [Cheng et al. \citeyear{cheng2008illuminance}; Fournier et al. \citeyear{fournier2008method}; Zhu et al. \citeyear{zhu2013analytical}; Poyet et al. \citeyear{poyet2016light}; Song et al. \citeyear{song2023analysis}] and optical storage [Lim et al. \citeyear{lim2023efficient}; Cheriere et al. \citeyear{cheriere2025holographic}]. As shown in Fig. \ref{fig:LP_propagate}, the light undergoes the kaleidoscopic guiding by TIR within a light pipe, thereby appears as if it is emanating from duplicated virtual light sources [Cheng et al. \citeyear{cheng2006optical}]. These virtual sources are flipped in the vertical or horizontal direction as TIR occurs. The light propagated from these wavefronts travels along the length of the light pipe and exits through the output aperture, where it is clipped.

The important aspect of this study is that if this propagation model is applied to an accurate wavefront reconstruction, the holographic image can be reconstructed after the light pipe with high quality suitable for display applications. 
This is plausible since the light pipe, under the use of fully coherent laser sources and well-controlled propagation geometry, effectively preserves both linear polarization and coherence properties necessary for holograms
%This is plausible since the light pipe preserves both linear polarization and coherence properties through propagation
[Roelandt et al. \citeyear{roelandt2013propagation}; Roelandt et al. \citeyear{roelandt2014influence}; Dickey et al. \citeyear{dickey2017laser}].
By utilizing this coherence-preserving behavior under such conditions, Caulfield et al. [\citeyear{caulfield1967light}] employed light emitted from the light pipe in interferometry and hologram recording. However, in this case, the FoV and three-dimensionality was limited. Lim et al. [\citeyear{lim2023efficient}] demonstrated efficient data transmission through a light pipe for optical storage by combining a ray-based localizer with a neural network decoder. While their method aims to recover spatially scrambled data for decoding process, it is based on the coarse ray-based spatial mapping and does not address precise wavefront propagation or holographic image reconstruction, which are essential for display applications.

One approach to model the kaleidoscopic wavefront guiding within a light pipe is to stitch the virtual wavefronts and then propagate the expanded wavefront. The total number of TIR $M$ can be calculated as follows:

\begin{equation}
\begin{aligned}
M &= \left\lceil \frac{l \cdot \tan\theta_{\text{max}}}{d} \right\rceil
\end{aligned}
\label{eq:maxFoldNum}
\end{equation}

Here, $d$ and $l$ denote the side length and propagation length of the light pipe. After stitching 2$M$ wavefronts for each axis, the wavefront $g$ at the end of the light pipe can be calculated using the angular spectrum method (ASM) as follows [Goodman \citeyear{goodman2005introduction}]:

\begin{equation}
\begin{aligned}
g(x, y; l) &= \mathscr{F}^{-1} \big[ \mathscr{F} \{g(x, y; 0)\} H(u, v; l)\big], \\
H(u, v; l) &=
\begin{cases} 
\exp \left[ j 2 \pi l
\sqrt{\lambda^{-2} - u^2 - v^2} \right], 
& \text{if } \sqrt{u^2 + v^2} < \lambda^{-1}, \\
0, & \text{otherwise}.
\end{cases}
\end{aligned}
\label{eq:asm}
\end{equation}

However, this simple approach has the drawback of high computational load due to the large size of the expanded domain. Instead, we calculate only the components of each virtual wavefront that reach the output aperture of the light pipe, as shown in Fig. \ref{fig:LP_propagate} (b). The output wavefront $g_{ij}(x, y; l)$ and its shifted ASM kernel for each virtual wavefront are given as follows [Matsushima \citeyear{matsushima2010shifted}]:

\begin{equation}
\begin{aligned}
g_{ij}(x, y; l) &= \mathscr{F}^{-1} \big[ \mathscr{F} \{g_{ij}(x, y; 0)\} H_{ij}(u, v; l)\big], \\
H_{ij}(u, v; l) &= H(u, v; l) \exp \big[ j 2 \pi (-x_{ij} u - y_{ij} v) \big] \\
&= \exp \big[ j 2 \pi \big( -x_{ij} u - y_{ij} v + l \sqrt{\lambda^{-2} - u^2 - v^2} \big) \big].
\end{aligned}
\label{eq:shiftedASM}
\end{equation}

Here, $i, j \in \mathbb{Z}$ is the location index ($-M \leq i, j \leq M$), and $(x_{ij}, y_{ij})$ is the center coordinate of each virtual wavefront (for the example of Fig. \ref{fig:LP_propagate} (b), $(x_{ij}, y_{ij}) = (d, d)$). To prevent aliasing, the kernel $H_{ij}$ is multiplied by a band limitation mask that satisfies the Nyquist sampling condition defined as follows [Matsushima \citeyear{matsushima2020introduction}]:

\begin{equation}
\begin{aligned}
\Delta u^{-1} &= 2d \\
&>  2 | x_{ij} - l\frac{u}{\sqrt{\lambda^{-2} - u^2 - v^2}} |, \\
\Delta v^{-1} &>  2 | y_{ij} - l\frac{v}{\sqrt{\lambda^{-2} - u^2 - v^2}} |. \\
\end{aligned}
\label{eq:shiftedASM_bandlimit}
\end{equation}

Note that $\Delta u^{-1}$ and $\Delta v^{-1}$ are twice the size of $d$ due to zero-padding, which is applied to prevent circular convolution. Finally, the wavefront $g_{LP}(x, y; l)$ at the output aperture of the light pipe can be obtained by summing $g_{ij}(x, y; l)$ as follows:

\begin{equation}
\begin{aligned}
g_{LP}(x, y; l) &= \sum_{i, j = -M}^{M} \big( \mathscr{F}^{-1} \big[ \mathscr{F} \{g_{ij}(x, y; 0)\} H_{ij}(u, v; l) \chi_{ij} \big] \big),
\end{aligned}
\label{eq:LP_final}
\end{equation}

where $\chi_{ij}$ is the band limitation mask of Eq. \ref{eq:shiftedASM_bandlimit}.

\subsection{Hologram rendering using light pipe model} \label{optimization}
%% 위에서 설명한 model 써서 홀로그램 최적화 하는 파이프라인 설명
%% 뒤에 application 부분에서 CITL 이나 converging 렌즈 포함해서 설명해야 돼서
%% 여기서는 pseudo code 써서 해야할듯.
%% 실제 셋업 설명, CITL 썼고, converging lens 로 dc filter 흐리게 한 스킴 설명,
%% 전체 최적화 스킴 그림으로 설명

The SLM used in the experiment is a reflective, phase-only modulator. Considering the complexity of the light pipe model, phase profile optimization is essential for rendering high-quality holograms. For a 2D image, the intensity reconstructed at the target distance z is given as follows:

\begin{equation}
\begin{aligned}
I &= \left| \mathscr{F} \left( g_{LP} \exp \left[ -j \frac{\pi}{\lambda z} \left( x^2 + y^2 \right) \right] \right) \right|^2
\end{aligned}
\label{eq:intensity}
\end{equation}

Consequently, the optimization of the SLM phase profile is performed to minimize the following MSE loss:

\begin{equation}
\begin{aligned}
\mathcal{L} &= \text{MSE} \big(s \cdot I, I_{\text{target}}\big)
\end{aligned}
\label{eq:loss_function}
\end{equation}

Here, s represents the energy coefficient. This optimization process is applied to multiple SLM profiles for temporal multiplexing and thereby reducing the speckle noise. The overall optimization process is visualized in Fig. \ref{fig:optimization}.

Theoretically, the rendered hologram should exactly match the target image. However, real optical setups inevitably introduce optical aberrations and distortions. To address this, a camera-in-the-loop (CITL) process can be adopted. CITL incorporates the actual captured holographic results into the optimization process [Peng et al. \citeyear{peng2020neural}; Chakravarthula et al. \citeyear{chakravarthula2020learned}; Choi et al. \citeyear{choi2022time}]. CITL operates under the assumption that the holographic images rendered through simulation are highly similar to the captured images. To account for distortions introduced by the optical setup, image rectification and homography transformation were applied to the captured and target images. The loss between the captured intensity and the target is then calculated to update the phase profile. As a result, the captured holographic image directly matches the original target image without additional geometric transformations.

%% ===============================================================================

\section{Holographic Display Experiments} \label{Experiments}
%% 수식들 용어 통일하기
\subsection{Holographic image result} \label{Benchtop prototype}

% \begin{figure*}[!t]
% \centering
%  \includegraphics[width=\linewidth]{Figure_lightPipe_Result_3D.pdf}
% \caption{Captured 3D holographic images. The focusing depth is written in diopters. SLM profiles of 10 temporal multiplexing are optimized based on the focal stack.}
% \label{fig:lightPipeResult3D}
% \end{figure*}

%% 위에 설명에서는 그냥 LP model 써야 올바른 이미지가 나오는 것을 알 수 있다! 정도로 써주고, 구체적인 내용은 본문으로 뺴자. 대신 실험 했던 방식 (monochrome 으로 찍고 RGB loss 최소가 되도록 색깔 합성했다, CITL 썼다, distortion corrected 됐다 이런걸 캡션에 써주자.)

We experimentally reconstruct holographic images using the proposed hologram rendering process of the light pipe propagation model. This experiment assumes a scenario where a user observes a holographic image through a thin image combiner, such as a light guide or waveguide, and accordingly renders a Fourier hologram.
As observed in the experimental results of Jang et al. [\citeyear{jang2024waveguide}], conventional Fourier holograms show a bright spot at the center of the image due to the DC noise of the SLM. To eliminate this issue while maintaining the quality of the Fourier hologram, the method proposed by Cho et al. [\citeyear{cho2018dc}] can be utilized. Cho et al. [\citeyear{cho2018dc}] introduced a converging wavefront to the SLM, separating the focusing plane of the DC noise from the holographic image plane. They further incorporated this converging phase profile into the SLM optimization process to generate DC-free holographic images. Likewise, we implement this approach by placing a converging lens to the collimated input light. The converging profile at the SLM plane is incorporated into the CITL optimization process to optimize the phase profile. The optimization and capturing process are performed independently for each color channel and digitally combined to form the final result.

\begin{figure}[!t]
\centering
 \includegraphics[width=\linewidth]{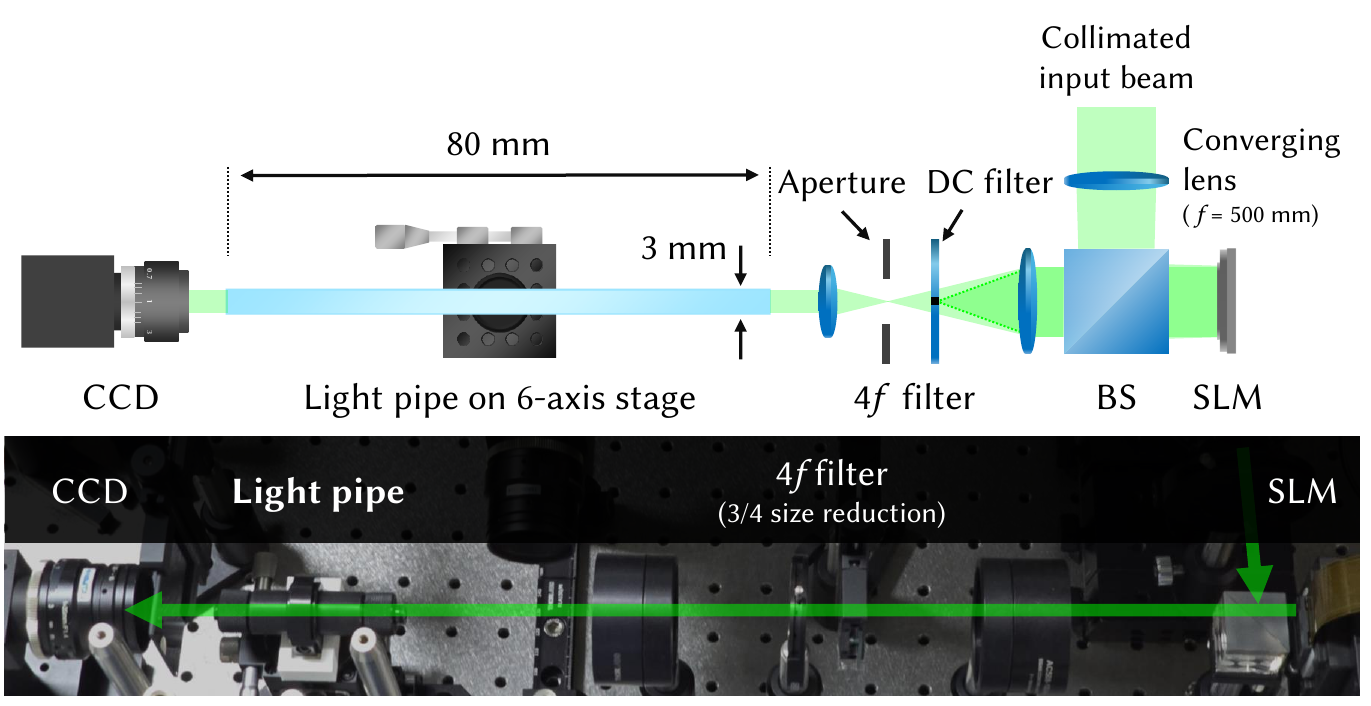}
\caption{(Upper) Schematic diagram and (lower) experimental setup of the proposed light pipe system. Due to the converging input beam to the SLM, the DC filter is installed away from the 4\textit{f} filter's focal plane. The wavefront size is reduced to match the light pipe's cross-section, and the holographic image is observed after the output aperture of the light pipe.}
\label{fig:schematic}
\end{figure}

\begin{figure*}[!t]
\centering
 \includegraphics[width=\linewidth]{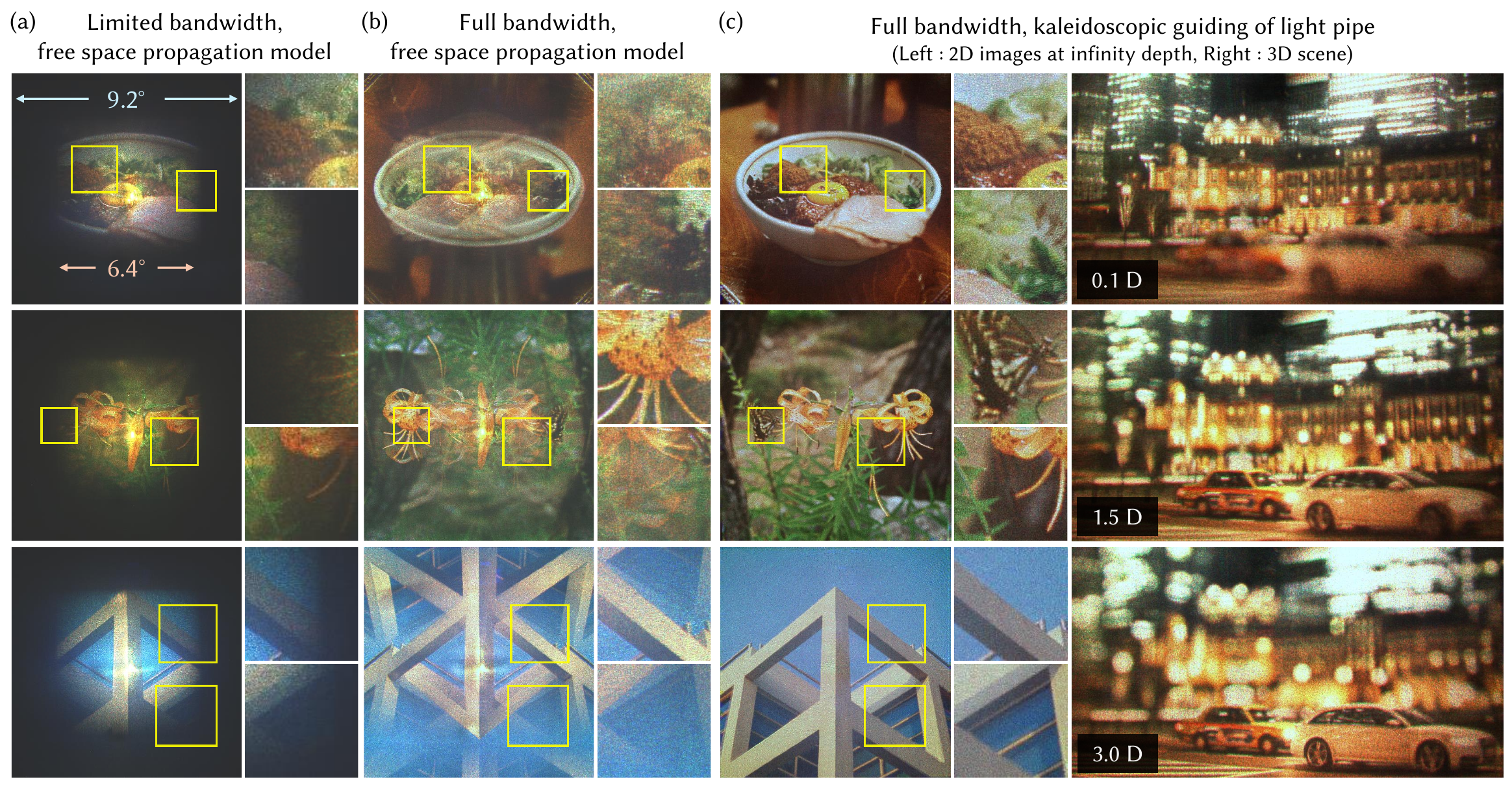}
\caption{Captured holographic images by rendering through different propagation model within the limited volume of a light pipe. 2D images are reconstructed at infinity depth, and the temporal multiplexing of 10 is applied to reduce the speckle noise. The focusing depth for 3D scene is written in diopters. The complete, correct image becomes observable with the full bandwidth when the proposed kaleidoscopic guiding model is integrated into phase optimization.}
\label{fig:lightPipeResult}
\end{figure*}

\begin{figure*}[!t]
\centering
 \includegraphics[width=\linewidth]{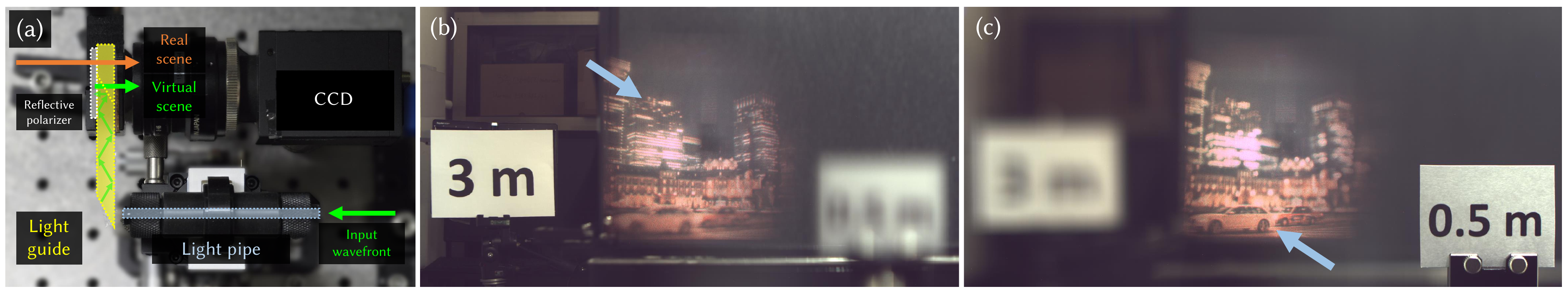}
\caption{(a) Experimental setup for AR prototype. (b-c) Captured AR scene of different depths. The wavefront is transferred to the light guide through kaleidoscopic guiding and propagated to CCD camera through partial reflective mirror and reflective polarizer. White balance is adjusted for the real scene, and color channels are digitally combined after capturing independently.}
\label{fig:arscene}
\end{figure*}

Figure \ref{fig:schematic} shows the schematic diagram and the physical setup of the experimental prototype. The laser wavelengths for red, green, and blue channels are 638 nm, 520 nm, and 450 nm, respectively. The light pipe (custom-fabricated by IRD Glass) used in the setup is made of N-BK7 optical glass and features high surface quality with a scratch/dig grade of 10/5 on both ends. The light pipe is a square type with $d=$ 3 mm. $l$ is set to 80 mm considering the typical length of eyeglass temples. This light pipe is installed on the 6-axis stage for further analysis (see Section \ref{misalignment_analysis}).

Light from the RGB laser is first collimated and then propagates through a lens with a focal length of 500 mm, resulting in a converging beam incident on the SLM. The SLM used in the experiment is the HOLOEYE GAEA-2.1 model with a sampling pitch of $p_{SLM}$ = 3.74 $\mu$m. The sampling number used for the experiment is $N$ =1000. The modulated wavefront passes through a 4\textit{f} filter, where the DC noise and high-order terms are filtered out before entering the light pipe. To match the wavefront size to the light pipe's cross-section, the 4\textit{f} filter is constructed with a 3/4 reduction ratio. The wavefront entering the light pipe propagates to the opposite end through TIR, and images at arbitrary depths are observed using a CCD camera.

Figure \ref{fig:lightPipeResult} presents the captured holographic images based on the propagation model used for hologram rendering. Images are reconstructed at the infinity. The first column depicts the images that can be observed when assuming a non-reflecting hollow pipe, where light propagates within the volume of the light pipe without TIR. While the full angular bandwidth of the SLM provides the FoV of 9.2$^{\circ}$, the effective FoV is limited to 6.4$^{\circ}$ due to the constrained form factor. The second column illustrates the kaleidoscope effect that occurs when the SLM profile is optimized using a free-space propagation model. Although the full angular bandwidth is preserved by TIR, the image becomes inverted and overlapped, making the original scene unrecognizable.

In contrast, the third and fourth column show the holographic images rendered based on the kaleidoscopic guiding model of the light pipe (Section \ref{Kaleidoscope}). The full bandwidth of the SLM is successfully transferred even with the limited form factor of the light pipe, and sharp images can be observed across the entire FoV. This bandwidth conservation further enables the realistic depth cues and blur effects for 3D scenes. These results validate that the proposed scheme maximizes the advantages of holographic displays while effectively decoupling the light engine from the image combiner.

\begin{figure*}[!t]
\centering
 \includegraphics[width=\linewidth]{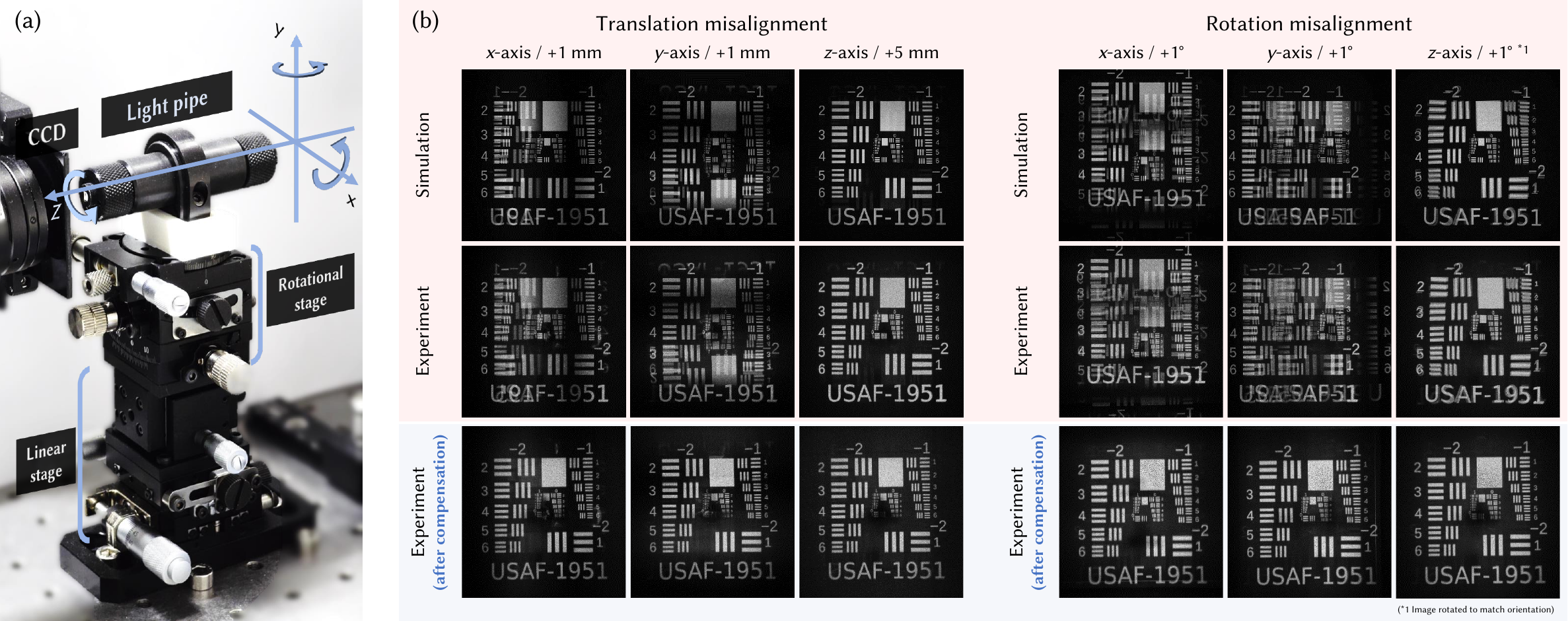}
\caption{Simulation and experimental results of mechanical misalignment and its compensation (USAF 1951 resolution chart, MIL-STD-150A). The brightness is adjusted to match each other, and images are converted from RGB color to monochrome. (a) Configuration of the 6-axis stage to translate and rotate the light pipe. (b) The simulated (top row) and experimentally captured (center row) image results upon misalignment type. For simplicity, each misalignment exists along a single axis. The simulated results show highly matching results from the captured ones. (Bottom row) Experimentally captured holographic image after the compensation of misalignment. The results validate that the propagation model of the light pipe is well-proposed.}
\label{fig:misalignment}
\end{figure*}

\subsection{AR scene result} \label{AR scene}

Based on the results presented so far, an NED prototype for AR is developed. The custom-fabricated N-BK7 light guide is used as an image combiner, positioned after the output aperture of the light pipe. Figure \ref{fig:arscene} shows the AR experimental setup and the captured AR scene. The thickness and hypotenuse angle of the light guide is 7 mm and 30$^{\circ}$, respectively. The wavefront entering the light guide undergoes internal TIR after reflecting from the slanted mirror surface. It is then directed toward the user's eye via a half-mirror (30$^{\circ}$ slanted) and a reflective linear polarizer. Simultaneously, light from the real scene passes through the light guide and reaches the user's eye. The results demonstrate that the 3D holographic image is successfully reconstructed, providing accurate depth perception. Moreover, the holographic image can be observed across the FoV corresponding to the full bandwidth of the wavefront which could be confirmed from Fig. \ref{fig:lightPipeResult}.

In conclusion, the kaleidoscopic guiding method using a light pipe is experimentally validated by developing an AR NED prototype. This approach successfully decouples the light engine from the image combiner while preserving the full bandwidth, and leave the transparent image combiner only for obstruction-free view. Although the prototype in this paper utilizes slightly thick light guide for the image combiner, further improved system with lightweight, smaller form factor can be achieved by using waveguide.

%% ===============================================================================

\section{Misalignment Analysis} \label{misalignment_analysis}
%% misalignment 발생했을 때 어떻게 보이는지, 어떻게 보상하는지 설명

We now further simulate the holographic images observed when misalignment occurs in the light pipe, and present a compensation method to reconstruct correct images. Since the wavefront modulation is a pixel-wise operation, holographic displays are inherently sensitive to mechanical misalignment. This issue becomes especially prominent when applied to holographic NEDs, as external forces in daily use can easily induce misalignment. Moreover, in the proposed scheme which models TIR components and incorporates them into phase optimization, such misalignment creates an even greater challenge and aggravates its impact on image quality.

We analyze mechanical misalignment by dividing it into six types: three directions of linear translation and rotations about three axes. Linear translations can be easily simulated by shifting and cropping the wavefront incident on the light pipe. Likewise, for rotation misalignment, we apply a rotational transform to the input wavefront so that the proposed kaleidoscopic guiding model computes the virtual wavefronts accordingly. The rotational transform is based on the delicate method proposed by Matsushima [\citeyear{matsushima2020introduction}].

Figure \ref{fig:misalignment} (a) shows the installed 6-axis stage on the light pipe for precise misalignment control. For rotation misalignment, both the simulation and experiment assume that the light pipe is rotated about the surface where the wavefront enters. Since the 6-axis stage is positioned at the center of the light pipe in the current setup, linear translation is applied after rotation to align the rotational axis with the input surface.

Figure \ref{fig:misalignment} (b) presents the effects of misalignment along each axis and the corresponding experimental results after compensation. It can be observed that the misalignment simulations based on the proposed model closely match the experimental results. While the outcomes of translation and rotation misalignment may appear similar, some distinctions exist.
In the case of translation, duplicated images prominently appear on one side. This phenomenon occurs because the cropped wavefront induces a loss of information on one side, and the shifted wavefront creates excessive TIR on the other. As a result, duplicated images are primarily observable in the high-frequency components (the outer regions of the Fourier hologram image), where a higher number of TIR occurs. Such information loss and variations in the number of TIRs predominantly happen by horizontal and vertical translations, whereas misalignment along the optical axis ($z$ axis) is comparatively robust. As shown in Fig. \ref{fig:misalignment} (b), even with a $z$ axis translation misalignment of 5 mm, the image changes are less significant compared to 1 mm misalignment along the $x$ and $y$ axes.

On the other hand, the entire wavefront rotates relative to the light pipe in the rotation misalignment. This results in replicated images being observed across all regions of the Fourier hologram, regardless of the original angular components. Additionally, since the propagation direction of the wavefront itself changes, the overall image is reconstructed to the shifted direction of rotation. Similar to translation, the image quality is comparatively more robust to rotation around the $z$ axis at the same rotation angles since the duplicated image simply rotates in its place.

The bottom row of Fig. \ref{fig:misalignment} (b) shows the holographic images captured after incorporating the misalignment into the SLM profile optimization. Note that this optimization is performed in a simulation environment rather than using CITL, and then captured after uploading the optimized SLM profile. In all types of misalignment, the correct image is successfully reconstructed. The compensation method is effective since the light pipe preserves linear polarization of the wavefront even after rotation [Sun et al. \citeyear{sun2010speckle}; Zhao et al. \citeyear{zhao2014influence}]. These results demonstrate that the proposed light pipe guiding model and optimization process are well-defined and effective. Furthermore, the ability to accurately compensate for mechanical misalignment validates the potential of the proposed approach for practical applications in holographic NEDs.

%% ===============================================================================

\section{Discussion} \label{Discussion}

\subsection{Advantages} \label{advantages}

%% 수식들 용어 통일하기
The proposed guiding approach is expected to offer greater advantages as improvements in SLM manufacturing increase the diffraction angle. Currently, the sampling pitch of SLMs remains at a few micrometer scale, which limits the achievable diffraction angle and field of view. Recently, there has been active research on the implementation of SLMs with a sampling pitch as small as 1 $\mu$m [Isomae et al. \citeyear{isomae2017design}; Choi et al. \citeyear{choi2019evolution}; Kim et al. \citeyear{kim2019crafting}; Yang et al. \citeyear{yang2023review}]. If the sampling pitch is reduced to the sub-wavelength scale, it would become possible to realize a holographic light engine with a large dispersion angle.

When the light with incidence angle $\theta$ enters the light pipe, the refracted angle $\gamma$ should fulfill the following Snell's law to be guided by TIR:

\begin{equation}
\begin{aligned}
\sin\theta &= n \sin\gamma, \\
\frac{\pi}{2} - \gamma &> \sin^{-1} \left( \frac{1}{n} \right).
\end{aligned}
\label{eq:snell_condition}
\end{equation}

Consequently, $\theta$ that can be guided through a light pipe with a refractive index $n$ is given as follows:

\begin{equation}
\begin{aligned}
\sin\theta &< \sqrt{n^2 - 1}.
\end{aligned}
\label{eq:sin_condition}
\end{equation}

In other words, a light pipe made of a medium with a refractive index $n > \sqrt{2}$ can guide the entire bandwidth via TIR. The N-BK7 material used in this study has a refractive index higher than $\sqrt{2}$ across the visible spectrum, ensuring that the proposed system remains applicable even if future advancements in SLM technology lead to increased diffraction angles.

\begin{figure}[!t]
\centering
 \includegraphics[width=\linewidth]{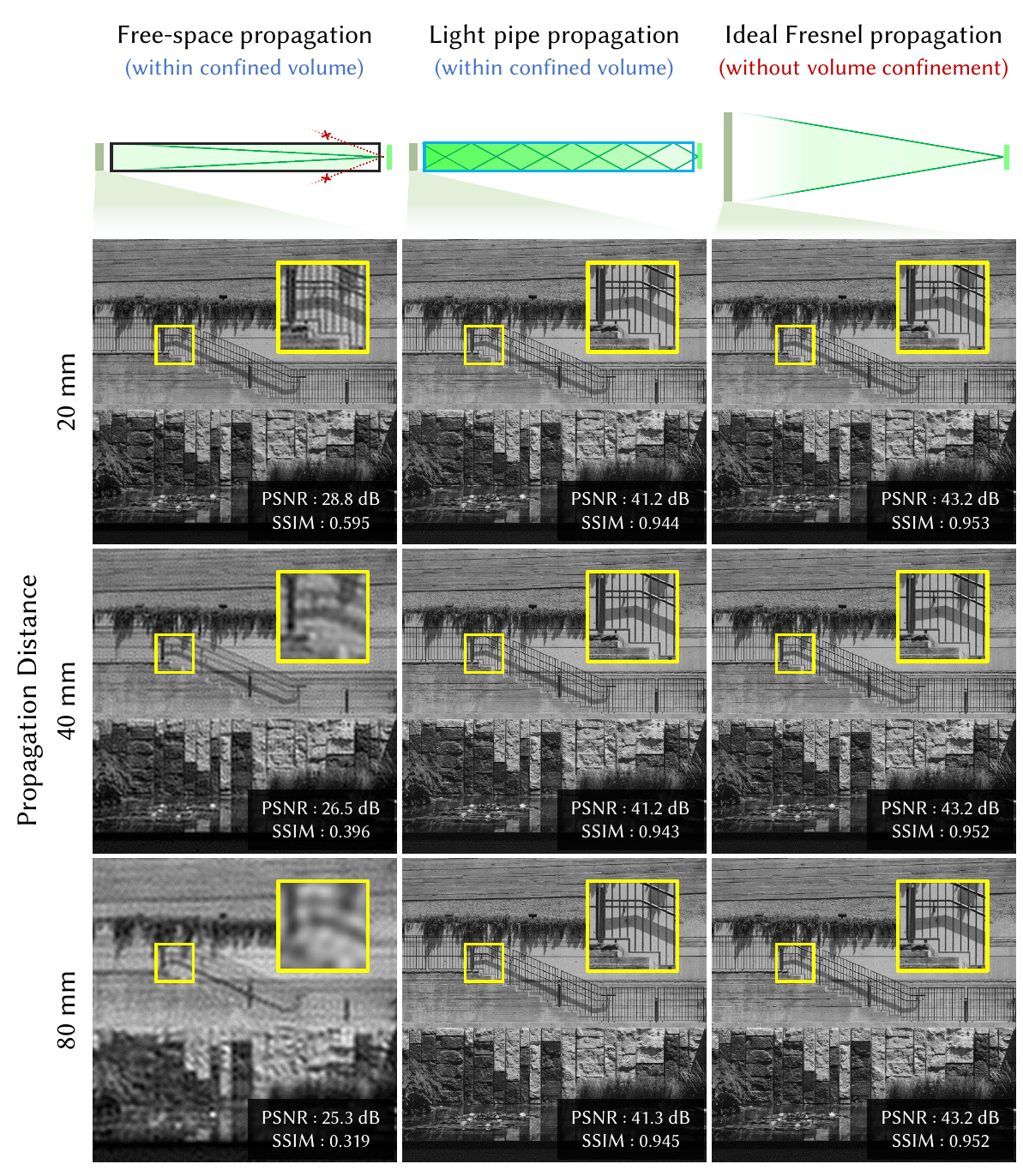}
\caption{Optimized Fresnel hologram by free-space propagation and the proposed light pipe system. Sampling pitch of SLM is 1 $\mu$m, and the wavelength is 450 nm. Temporal multiplexing of 2 is applied. For propagation within a confined volume (left and center columns), the sampling window is limited to 1 mm $\times$ 1 mm. The ideal Fresnel propagation (right column) assumes an unbounded volume where the physical sampling window expands with increasing propagation distance. The proposed system has advantages in reconstructing holographic images with high quality for all propagation distances within the constrained volume.}
\label{fig:freq_loss}
\end{figure}

Furthermore, as the diffraction angle increases, the loss of high-frequency components in free-space propagation within a confined volume becomes significant. This issue manifests as an FoV limitation in Fourier holograms (as shown in Fig. \ref{fig:lightPipeResult}) and as spatial frequency loss in Fresnel holograms. Figure \ref{fig:freq_loss} illustrates Fresnel holographic images optimized using both free-space propagation and the light pipe guiding method, simulated for an SLM with a sampling pitch of 1 $\mu$m. In the volume confinement cases, the sampling window is restricted to 1 mm $\times$ 1 mm. As the propagation distance increases, free-space propagation exhibits a noticeable loss of high-frequency components that reach the same sampling window, resulting in a decrease in the spatial frequency of the image. In contrast, the light pipe guiding method preserves higher spatial frequencies via TIR regardless of its length, comparable to the result of ideal Fresnel propagation without volume confinement. This indicates that the proposed method is particularly advantageous for delivering high-quality holographic images when advanced SLMs are applied to AR NEDs in the future.

\subsection{Challenges and future works} \label{challenges}
% \subsubsection*{Model mismatch analysis and its compensation} \label{misalignment}
%% Fiber와의 경계가 아직은 좀 모호함을 인정.
%% 레이저 파워 쎄지면 파장 broad 해져서 복제텀? 이거 근데 생기긴하는데 시뮬상으로는 안생기는게 너무 미스테리긴 함.

\subsubsection*{\indent Alignment sensitivity}
Although the proposed method can resolve the image degradation induced by the misalignment, use of light pipe still makes the overall system alignment-sensitive. As can be seen in Fig. \ref{fig:misalignment}, the duplicated images are mainly induced by the misalignment, which could be more irritating to the user than simple optical aberrations. When utilized for the compact hand-held AR glasses, it is difficult to expect the aspect of misalignment in a real time. The future study about thorough misalignment measurement and compensation by real-time feedback is needed, which will potentially increase the system complexity.

\begin{figure}[!t]
\centering
 \includegraphics[width=\linewidth]{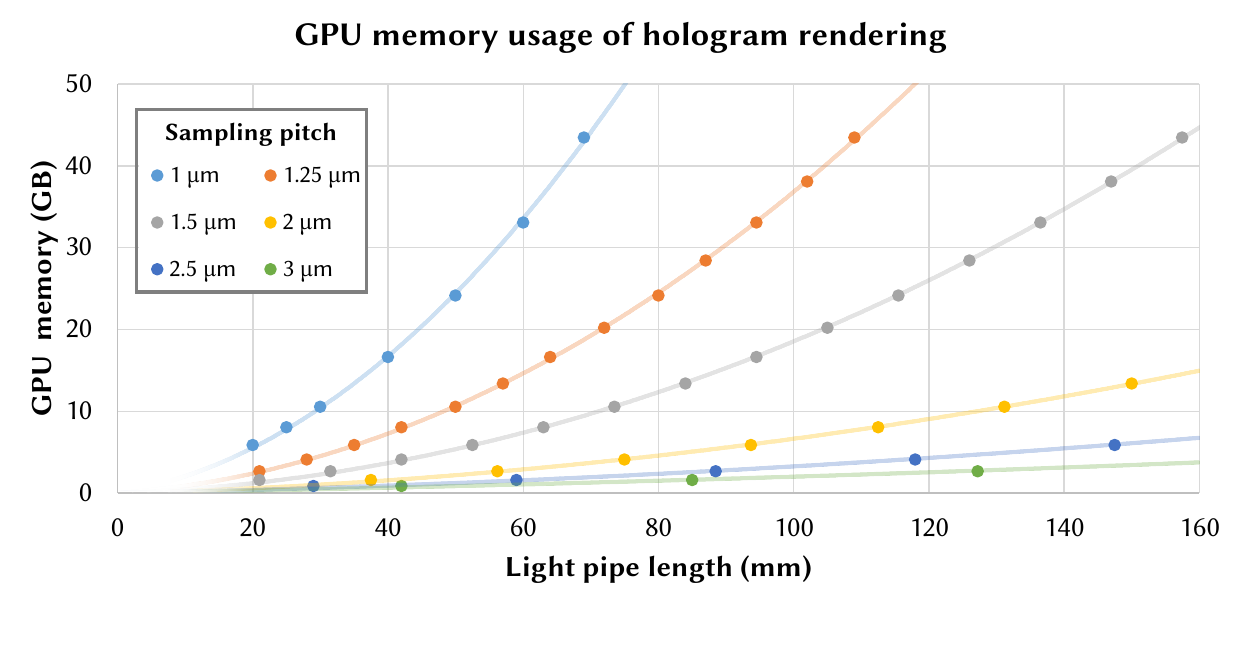}
\caption{GPU memory usage of CGH rendering through light pipe's kaleidoscopic guiding. To analyze the worst-case scenario, the wavelength is set to 638 nm (red) which results in the highest number of TIRs due to its large diffraction angle. The sampling number is 1000 for all cases, and temporal multiplexing of 10 is applied.}
\label{fig:gpu_memory}
\end{figure}

\subsubsection*{\indent Computation efficiency}
The number of kernels required for the shifted ASM increases quadratically with the number of TIR. In this experiment, 25 shifted ASM kernels were required for red light, and this number and subsequent computational load increase as the light pipe gets longer.
Figure \ref{fig:gpu_memory} illustrates the GPU load of CGH rendering through the light pipe's kaleidoscopic guiding at a wavelength of 638 nm. The GPU memory usage exhibits a quadratic increase with respect to the length of the light pipe. For a sampling pitch of 1 $\mu$m, the rendering process occupies 43.2 GB of GPU memory to simulate 15 TIRs within 1 mm-thick light pipe, corresponding to a propagation length of 69.5 mm—comparable to the typical length of eyeglass temples. This computational demand is within the capability of existing hardware (RTX A6000 GPU). Given that a 1 $\mu$m sampling pitch enables a Fourier hologram with approximately 37° horizontal and 51° diagonal FoV at 638 nm, Fig. \ref{fig:gpu_memory} demonstrates the feasibility of the proposed method for AR NEDs with wider FoV when 1 $\mu$m SLMs become commercially available.

However, further reduction of sampling pitch and increase in light pipe length would require substantial computational load. Moreover, the rendering process is not conducted in real time. To address this limitation, further research is required to develop more efficient computation, since each shifted ASM kernel inherits sparsity of values caused by the band limitation mask. Exploring a trade-off between computational complexity and propagation model's accuracy may also improve the rendering efficiency at the cost of holographic image fidelity, as suggested by Lim et al. [\citeyear{lim2023efficient}] through coarse ray-based localizer.
Additionally, incorporating neural network-based propagation model training into the optimization process could potentially realize the higher-quality holographic images in faster speed [Choi et al. \citeyear{choi2021neural}].

\subsubsection*{\indent Further miniaturization}
The current thickness of the light pipe is 3 mm, which was custom-fabricated considering the manufacturing difficulty. For a more compact system, a thinner light pipe of 1 mm or less would be desirable. However, as the thickness decreases, the manufacturing complexity increases significantly. Additionally, adding a light pipe to the device inevitably results in a slight increase in the overall weight of the system. Therefore, further research into lightweight and optically uniform materials for light pipes is necessary for practical implementation. Moreover, if the thickness of the light pipe is continuously reduced to the order of tens of micrometers, geometric interpretation may no longer properly describe the behavior of the light. Instead, it would be necessary to account for propagation modes, similar to the optical fibers. Identifying the critical thickness at which this transition occurs and determining the feasibility of experimental implementation are important areas for future research.

%% ===============================================================================

\section{Conclusion} \label{Conclusion}

In this work, we proposed a kaleidoscopic guiding method for delivering full-bandwidth wavefronts with a light pipe. Using this approach, we repositioned the light engine away from the image combiner and suggested a versatility in near-eye display design. We introduced an SLM phase optimization technique using the light pipe and rendered high-quality 3D holographic images. We also simulated the misalignments of the light pipe and proposed a robust CGH rendering method that accounts for these misalignments. Furthermore, we implemented a setup using a custom-fabricated light pipe designed for practical AR displays, and experimentally verified the bandwidth preservation and high-quality 3D holographic images even within highly limited volumes. Although there remains room for improvement in the proposed system, we expect that the proposed methods can be applied to more practical and versatile AR or holographic display systems in the future.

%% ===============================================================================

\begin{acks}
This work was supported by the BK21 FOUR program of the Education and Research Program for Future ICT Pioneers, Seoul National University in 2025.
\end{acks}

\bibliographystyle{ACM-Reference-Format}
\bibliography{sample-base}

\end{document}